\documentstyle[12pt]{article}
\begin{document}   

\newcommand {\be} {\begin{equation}}
\newcommand {\bea} {\begin{eqnarray} \nonumber } 
\newcommand {\ee} {\end{equation}} 
\newcommand {\eea} {\end{eqnarray}}
 \newcommand {\eps} {\epsilon}
 \newcommand {\s} {\sigma} 
 \newcommand {\si} {\sigma}
\newcommand {\de} {\delta}
 \newcommand {\g} {\gamma} 
\newcommand {\la} {\lambda}
\newcommand {\La} {\Lambda}
\newcommand {\al} {\alpha}
\newcommand {\N} {{\cal N}}
\newcommand {\cP} {{\cal P}}
\newcommand {\NE} {\not=}
\newcommand {\ba} {\overline} 
\newcommand {\lan} {\langle}
\newcommand {\ran} {\rangle}
\newcommand {\bi} {\bibitem}
\newcommand {\Tr} {\mbox{Tr}}
\newcommand {\qu} {q_{1}}
\newcommand {\qd} {q_{2}}
\newcommand {\qt} {q_{3}}

\title{On the probabilistic formulation of the replica approach to spin glasses}

\author{  Giorgio Parisi \\ 
Dipartimento di Fisica,
Universit\`a {\sl La  Sapienza}\\ INFN Sezione di Roma I \\ Piazzale 
Aldo Moro 2 , Roma 00185, Italy}
\maketitle

\begin{abstract}
In this paper we review the predictions of the replica approach on the probability distribution of 
the overlaps among replicas and on the sample to sample fluctuations of this probability.  We stress 
the role of replica equivalence in obtaining  relations which do not depend on the form of 
replica symmetry breaking. A comparison is done with the results obtained with a different rigorous 
approach.  The role of the ultrametricity and of other algebraic properties in discussed. It is shown 
that the ultrametric solution can be obtained from a variational principle.
\end{abstract}
\section{Introduction} 

The mean field theory for spin glasses has been originally derived by finding the solution of the 
Sherrington Kirkpatrick model \cite{MPV,PB2}.  The replica method was used in the first derivation.  
Probabilistic tools (based the cavity method) can be used instead of the replica approach, the 
formalism being more esplicite at the price of being more heavy \cite{MPV,PB2,PARJSP}.

The probabilistic method and the replica approach are conceptually 
equivalent.  In both cases the ultrametric structure of the states was assumed from the beginning.  
In order to present a definitive proof of the proposed solution, one should prove that  the 
ultrametric Ansatz is the unique stable solution of the mean field equations, or, if there are more 
stable solutions, one should show that the ultrametric one has be chosen.  This proof is lacking 
because of the difficulties in considering a generic non ultrametric Ansatz. 

 The aim of this note 
is to perform a small step in this direction by combining probabilistic methods and the replica 
techniques and by showing that the usual ultrametric solution can be selected by using a variational 
principle.  In section II and III we briefly review to replica technique and the meaning of 
spontaneously broken replica symmetry.  In section IV we spell out some of the consequences of 
replica equivalence and its connections with rigorous results.  In section V we study ultrametricity 
and in section VI we show how the ultrametric solution may be singled out  using an appropriate 
variational principle.  In section VII we mention the existence of two extra symmetry, translational 
invariance in replica space and separability, which strongly constrain the probability distribution.

\section{The real replica approach}

In this paper many of the considerations will not depend on the detailed form of the interaction.  We 
will study a system whose dynamical variables are Ising spins ($\s_i=\pm 1$, $i=1,N$).  The total 
number of degrees of freedom is $N$ which is assumed to be quite large but finite.  The system is at 
thermal equilibrium at a temperature $T=\beta^{-1}$.

The Hamiltonian will depend on some control variables $J$.  A standard procedure consists in 
considering an ensemble of systems, which differ by the value of the control parameters $J$ and to 
compute the average over $J$ \cite{EA}.  In many cases, i.e.  spin glasses, the number of parameters 
$J$ is proportional to $N$; an other possibility consists in considering a fixed value of $J$ and 
averaging over $N$ inside a given interval (e.g.  $N_{0}-\Delta<N<N_{0}+\Delta$ with $\Delta$ large
\cite{NS}).  In the following the average over the $J$ variables (or over $N$) will be denoted by a 
bar.

Let us firstly study what happens at a given value of $J$ (or $N$).
It is useful to consider $m$ 
replicas of the same system ($m$ being arbitrary, may be infinite) and to define the overlaps
\be
q_{s,t}=\frac{\sum_{i=1,N} \s_i^s \s_i^t}{N},
\ee
where $s$ and $t$ denote the replica.  

Here the number of replicas $m$ is a positive integer, which 
remain fixed and will {\sl not } go to zero at the end of the computation; this number should not be 
confused with $n$, the number of replicas that are used to compute the statistical expectation 
values, which eventually is taken to be equal to zero.  These replicas are just $m$ copies of the 
same system; their properties are in principle experimentally observables (they can be studied with 
numerical simulations) and some times they are called {\it real} replicas \cite{REAL}.

We want to explore the possibility that the quantity $q$ still fluctuates also for very large 
(obviously {\sl finite}) systems.  For each choice of the parameters $J$ we can study 
the probability distributions of the overlaps.  
We can define various functions, the simplest one $P_J(q)$ defined
\be
\int dq P_J(q) f(q)\equiv <f(q_{1,2})>=\ba{<f(q_{s,t})>_{J}}.
\ee
where $s$ and $t$ are arbitrary different replicas (all replicas are intrinsically equal!).  In other 
words $P_J(q)$ is the probability that two different equilibrium configurations have overlap $q$.

In we consider more overlaps together, we have to 
introduce other probability distributions: e.g.  $P_J^{12,23,31}(q_{1,2},q_{2,3},q_{3,1})$ 
defined as:
\be
\int d q_{1,2}dq_{2,3}dq_{3,1} P_J^{12,23,31}(q_{1,2},q_{2,3},q_{3,1})
 f(q_{1,2},q_{2,3},q_{3,1})=
<f(q_{1,2},q_{2,3},q_{3,1})>_{J}.
\ee

There are other probability distributions which can be trivially obtained as functions of  others.  
An example is given by $P_J^{12,34}(q_{1,2},q_{3,4})$ defined as
\be
\int dq_{1,2}dq_{3,4} P_J^{12,34}(q_{1,2},q_{3,4}) f(q_{1,2},q_{3,4})=
<f(q_{1,2},q_{3,4})>_{J}.
\ee
It is obvious that all the configurations $\s_i^s$ have the same probability independently from
the value of $s$, so that the probability distribution of the overlaps will be symmetric under the 
exchange of the replica indices.  We thus have
\be
P_J(q_{1,2})=P_J(q_{3,4}),\ \ \ 
P_J^{12,34}(q_{1,2},q_{3,4})=P_J(q_{1,2})P_J(q_{3,4}).
\ee

By increasing the number of replicas we find that more and more probability functions are needed. Each 
system (for given $J$ and $N$) may have the functions $P$ which differs from those of an other 
system.  Indeed at low temperature, where these functions are not given by a single $\delta$ 
function, they change of a quantity of order 1 when the total Hamiltonian changes of a quantity of 
order 1 (e.g.  when we change the form of the Hamiltonian in one point).  

The actual 
form of the functions $P$ for a given systems has the disappointing property of not being 
thermodynamical stable, i.e.  it changes completely under any perturbation.  As we shall see later, 
this is the price to pay in order to have that the probability distribution of the functions $P$.  
i.e.  $\cP [P]$, is stable with respect to a wide class of perturbations \cite{GU,AC}.

In order to get a feeling of the meaning of the fluctuations of the overlap, let us suppose that the 
Boltzmann average for {\sl finite} but large $N$ can be decomposed into different pure states or 
valleys:
\be
<\cdot>=\sum_{\alpha}w_{\alpha}<\cdot>_{\alpha}\label{DECO1}
\ee
and the expectation value in each of the pure state $\alpha$ satisfy the cluster property for the 
correlations functions, i.e.  intensive quantities do not fluctuate.

The name {\sl pure state} is abusive from a strict point of view: intensive quantities do always 
have fluctuations of order $N^{-1}$ and for $N$ fixed (albeit large but finite) we do not have any 
criterion (apart common sense) to distinguish fluctuations which are of order 1 from those that are 
of order $N^{-1}$.

In the usual ordered case one can easily define the states for the actual 
infinite system.  In disordered systems  the limit $N\to \infty$ may be rather tricky and it is 
quite possible that the expectation values of local observables do not have a limit (i.e.  they 
oscillate) when $N\to \infty$.  Here we do not want to enter into rather sophisticated mathematical 
arguments and we limit ourselves to study finite and large systems and to perform the limit 
$N\to\infty$ only after the average over $J$ (or $N$).

 If we assume that equation (\ref{DECO1}) can be approximately written 
for also for finite system, we can approximately write (using the asymptotic vanishing of the 
connected correlations within a pure state):
\bea
P_J(q)=\sum_{\al,\gamma}w_{\alpha}w_{\gamma} \delta(q^{\alpha,\gamma}-q),\\
P_J^{12,23,31}(q_{1,2},q_{2,3}q_{3,1})=\sum_{\al,\gamma,\beta} w_{\alpha}w_{\gamma}w_{\beta}
\delta(q_{\alpha,\gamma}-q_{1,2}) \delta(q_{\alpha,\beta}-q_{2,3}) \delta(q_{\beta,\gamma}-q_{3,1}).
\eea

In general these probabilities are $J$-dependent.  It is interesting to consider their average over 
the $J$'s, which for simplicity we will indicate removing the suffix $J$:
\begin{eqnarray} \nonumber
P(q_{1,2})=\ba{P_J(q_{1,2})},\ \ 
P^{12,23,31}(q_{1,2},q_{2,3},q_{3,1})=\ba{P_J^{12,23,31}(q_{1,2},q_{2,3},q_{3,1})}, \\ 
P^{12,34}(q_{1,2},q_{3,4})=\ba {P_J^{12,34}(q_{1,2},q_{3,4}}).
\end{eqnarray}

The number of independent probability distributions is infinite.  The replica method of next 
section consists in coding all this infinite set of functions into a $n
\times n$ matrix $Q$.  The value of $n$ is originally an integer.  At a later stage non integer 
values of $n$ are allowed and eventually the limit $n \to 0$ is performed.

\section{Breaking the replica symmetry}
The replica method uses a $n \times n$ matrix $Q$.  The free energy can be written as function of 
$Q$ in the limit $n \to 0$ and is invariant under the action of the permutation group of $n$ 
elements acting on indices of $Q$.  In this section we will study the relations among the properties 
of the matrix $Q$ and the probability distributions introduced in the previous section.

When replica symmetry is exact all the off diagonal elements of the matrix $Q$ are equal (the 
diagonal elements of $Q$ are equal to zero). When replica symmetry is spontaneously broken, we must 
consider a matrix $Q$ which is not invariant under the permutations of replicas.  If $\Pi$ is a 
permutation and $\Pi(a)$ is the value of the index $a$ after a permutation, the value of
\be
Q_{\Pi(a),\Pi(b)}\equiv Q^{\Pi}_{a,b},
\ee 
for fixed $a$ and $b$,  depends on the permutation $\Pi$.  We can thus introduce the function 
$P^R(q_{1,2})$ which is the probability distribution of the quantity $Q^{\Pi}_{1,2}$ averaged over  
all $n!$ permutations $\Pi$ with equal weight.  In other terms we define the function $P_R(q_{1,2})$ 
as follows;
\begin{equation}
{\sum_\Pi 
f(Q^{\Pi}_{1,2})
 \over n!}
\equiv
 \int dq_{1,2} P_R(q_{1,2})f(q_{1,2}),
\end{equation}
$f$ being an arbitrary test function.
In the same way we can define other functions, e.g. 
\be
{\sum_\Pi f(Q^{\Pi}_{1,2},Q^{\Pi}_{3,4})\over n!}\equiv
\int dq_{1,2} dq_{3,4} P_R^{12,34}(q_{1,2},q_{3,4})f(q_{1,2},q_{3,4}).
\ee
For each matrix $Q$ for a given $n$ we define a set of probability functions $P_{R}$, which are in 
one to one correspondence with the previous introduced functions
\footnote{If we are not interested to consider the overlap of a configuration with itself, that for 
Ising spin is 1, we can consider only the off diagonal terms of the matrix $Q$, which it is 
conventionally taken to be zero on the diagonal.}.

The main assumption of the replica method is 
that there exists a matrix $Q$ such that for a given value of $n$ (in particular for $n=0$) the 
$P_R$ functions defined in this way do coincide with the homologous probabilities $P$ previously 
defined with the real replicas.  The case $n=0$ seems rather strange; it is defined by considering 
the previous formulae for integer $n$ and perform the analytic continuation to $n=0$.  At the 
present moment it is not clear how much restrictive this hypothesis is.  The space of all possible 
matrix $Q$ analytically continued at $n=0$ is infinite dimensional and it is not easy to prove that 
a given set of $P$ functions cannot be obtained in this way.

In other words we have traded an infinite matrix $q_{a,b}$ (if $m=\infty$), which fluctuates also 
for very large systems, with a matrix $n \times n$ matrix $Q$ (which does not fluctuate) in the 
limit $n
\to 0$.  We have coded the infinite number of functions, which characterize the probability
distribution of $q$, in the structure of the matrix $Q$.  

As we said in the introduction the only well 
studied example of matrix $Q$ is such that the probability distribution of the overlaps is 
ultrametric, i.e. if $q_{1,2}>q_{1,3}$ and $P(q_{1,2},q_{2,3},q_{3,1})>0$ then $q_{2,3}=q_{1,3}$.  In 
this ultrametric scheme the relevant quantity is $P(q_{1,2})$ and (as we shall see later) all 
other high order functions can be obtained from it by algebraic manipulations
\cite{MPV,PB2,PARJSP}.

\section{Replica Equivalence}
\subsection{Introducing Replica Equivalence} 

Let us now consider a more general case in which ultrametricity is not assumed.  There are general 
arguments which imply that the form of the matrix $Q$ is not completely arbitrary \cite{MPV}.  It 
has been argued that also when replica symmetry is broken the matrix $Q$ must be such that no 
replica has different characteristics from the others.  Observables which involve only one replica 
must be replica symmetric.  In particular we must have that
\be
\sum_b f(Q_{a,b})\label{DEMO}
\ee
does not depend on $a$ \footnote{This is essentially equivalent to assume that each line is a 
permutation of other lines, which would be true for example in the case of Knuth's combinatorial 
matrices \cite {KU}.}.  This properties is called replica equivalence.  It is likely connected to 
the fact that the expectation values of intensive ($J$-independent) quantities do not change from 
one state to the other.

This hypothesis  implies automatically in the replica formalism that the free energy is 
finite.  Indeed in the replica approach the free energy can be written as function of the matrix 
$Q$ (see eq.  (\ref{FREE}).  If replica equivalence is correct, the term $1/n$ in the free energy 
cancels automatically with the sum over $n$.  Moreover adding the appropriate terms in the 
Hamiltonian we can generate an extra term proportional to
\be
{\sum_{a,b} f(Q_{a,b})\over n}.
\ee
This term is automatically finite if condition (\ref{DEMO}) is satisfied. 

 It interesting to note 
that this argument is quite similar to those which has been used by Guerra \cite{GU} in his 
rigourous derivation of some of the following equations.  The replica equivalence conditions is also 
deeply related to the stability conditions of Aizenman and Contucci \cite{AC}.

\subsection{Two overlaps}
Condition (\ref{DEMO}) has  interesting consequences.  In the simplest case we can consider 
the following equations:
\be
\sum_{c,d}Q_{a,c}^{k_{1}}Q_{b,d}^{k_{2}}=(\sum_{c,d}Q_{a,c}^{k_{1}})^{2}=\int 
dq_{1}P(q_{1})q_{1}^{k_{1}}\int dq_{2}P(q_{2})q_{2}^{k_{2}}.
\ee

The previous equation holds both for $a=b$ and for $a\ne b$.  Let us spell its consequences in the 
first case.  Its l.h.s can be written as
\be
\left(\sum_{c,d;c\ne d}+\sum_{c,d}\delta_{c,d}\right)\left(Q_{a,c}^{k_{1}}Q_{a,d}^{k_{2}}\right).
\ee
In the first sum there are $(n-2)(n-1)$ non zero terms ($Q_{a,a}=0$) and in the second sum there are 
$(n-1)$ non zero terms.  Each term of the first and second sum give a contribution equal 
respectively to
\be
\int dq_{1,2} dq_{1,3}P^{12,13}(q_{1,2},q_{1,3})q_{1,2}^{k_{1}}q_{1,3}^{k_{2}}\ , \  \ \ \
\int dq_{1,2} P(q_{1,2})q_{1,2}^{k_{1}+k_{2}}\ .
\ee
Putting everything together we obtain the relations
\be 
P^{12,13}(q_{1,2},q_{1,3})=\frac12 P(q_{1,2})P(q_{1,3})+
\frac12 P(q_{1,2}) \delta(q_{1,2}-q_{1,3}).\label{PRIMA}
\ee
A similar equation can be obtained if we consider the case $a\ne b$.  Indeed after a similar algebra 
and using the previous results we get
 
\be
P^{12,34}(q_{1,2},q_{3,4})=\frac23 P(q_{1,2})P(q_{3,4})+
\frac13 P(q_{1,2}) \delta(q_{1,2}-q_{3,4}).\label{RELA}
\ee

These relations have a peculiar standing because they have been proved by Guerra \cite{GU} under 
quite general conditions.  We can thus safely assume that they are correct, without any further 
reference to the replica method.

\subsection{Three overlaps}
Using the same method as before we obtain that replica equivalence implies the following relations:
\begin{eqnarray} 6 P^{12,13,14}(q_{1},q_{2},q_{3})=P(\qu)P(\qd)P(\qt)+ \nonumber \\
\delta(\qu-\qd)P(\qu)P(\qt)	+\delta(\qu-\qt)P(\qu)P(\qd)+\delta(\qd-\qt)P(\qu)P(\qd)+\\
 2\delta(\qu-\qd)\delta(\qd-\qt)P(\qu) \ .\nonumber
\end{eqnarray}
\begin{eqnarray} 
 6 P^{12,13,34}(q_{1},q_{2},q_{3})=2P^{12,23,31}(\qu,\qd,\qt)+P(\qu)P(\qd)P(\qt)+ \nonumber \\
\delta(\qu-\qd)P(\qu)P(\qt)	+\delta(\qd-\qt)P(\qu)P(\qd)+
\delta(\qu-\qd)\delta(\qd-\qt)P(\qu)\ ,\nonumber
\end{eqnarray}
\begin{eqnarray} 
12P^{12,34,15}(q_{1},q_{2},q_{3})=
2P^{12,23,31}(\qu,\qd,\qt)+3 P(\qu)P(\qd)P(\qt) +\nonumber \\
\delta(\qu-\qd)P(\qu)P(\qt)	+3\delta(\qu-\qt)P(\qu)P(\qd)+\delta(\qd-\qt)P(\qu)P(\qd)+
\\ 2\delta(\qu-\qd)\delta(\qd-\qt)P(\qu)\ ,
\end{eqnarray}
\begin{eqnarray} 
15P^{12,34,56}(\qu,\qd,\qt)= 2P^{12,23,31}(\qu,\qd,\qt)+5 P(\qu)P(\qd)P(\qt)+ \nonumber \\
+2\delta(\qu-\qd)P(\qu)P(\qt) +2\delta(\qu-\qt)P(\qu)P(\qd)+2\delta(\qd-\qt)P(\qu)P(\qd)+
\\ 2\delta(\qu-\qd)\delta(\qd-\qt)P(\qu).\nonumber
\end{eqnarray}
Some of these relations have been derived in \cite{AC} using a different technique.

It is remarkable 
that all the probability functions involving three overlaps can be computed from that with one 
overlap plus $P^{12,23,31}$.  This last function if quite important because it contains the main 
information on the validity of ultrametricity.

\subsection{Further consequences}

If we look in more details to replica equivalence there are further relations that must be satisfied.
For example we could impose that
\be
T_{a}=\sum_{b,c}q_{a,b}^{k_1}q_{b,c}^{k_2}q_{c,a}^{k_3}
\ee
does not depend on $a$.

For example we could spell the consequences of the condition
\be
{\sum_a T_a^2 \over n} =({\sum_a T_a \over n})^2
\ee
If we proceed as before we find that
\bea
6 P^{12,23,31,14,45,51}(q_1,q_2,q_3,q_4,q_5,q_6)
+P^{12,23,31}(q_1,q_2,q_3)\delta(q_4-q_1)\delta(q_5-q_2)\delta(q_6-q_3)=\\
P^{12,23,31}(q_1,q_2,q_3) P^{14,45,51}(q_4,q_5,q_6)+\\
+3(P^{12,23,31,24,41}(q_1,q_2,q_3,q_5,q_6)\delta(q_4-q_2)
+P^{12,23,31,14,43}(q_1,q_2,q_3,q_4,q_5)\delta(q_3-q_6). \nonumber
\eea
Many relations of this kind may be derived and it is quite likely that {\sl all} the equation that 
can be derived from replica equivalence can also proved using the method of \cite{GU,AC}.

\section{Ultrametricity}

The ultrametricity condition states that the function $P^{12,23,31}$ can be written as
\bea
P^{12,23,31}(q_{12},q_{23},q_{13})=\\
\theta(q_{12}-q_{13})\delta(q_{23}-q_{13})A(q_{12,}q_{13})
+\mbox{ 2 cyclic permutations } \\
+B(q_{12})\delta(q_{12}-q_{13})\delta(q_{23}-q_{13})\nonumber
\eea
It is already been shown \cite {IPR} that this condition and the previous relations, in particular 
eq.~(\ref{PRIMA}), implies that
\bea
A(q_{12},q_{1,3})=P(q_{12})P(q_{1,3}),\\
B(q_{12})=x(q_{12})P(q_{12})\label{83}
\eea
where 
\be x(q)=\int_{0}^{q} dq' P(q'), \ee
and we have assumed for simplicity that the support of the function $P(q)$ is within the interval 
[0-1]. 

 This relation has been proved \cite {IPR} by noticing that
\be
\int_{0}^{1} dq_{23} P^{12,23,31}(q_{12},q_{23},q_{13})=P^{12,13}(q_{12},q_{13}).
\ee
The r.h.s is given by eq. (\ref{PRIMA}) while the l.h.s.  is given by
\begin{equation}
A(q_{12},q_{13})+ \int{q_{12}} ^{1} dq' (A(q_{12},q') \delta(q_{12}-q_{13})
+ B(q_{12})\delta(q_{12}-q_{13})).
\end{equation}
We finally find that the usual result (\ref{83}) is correct.

In the same way it can be proved that in the ultrametric case also the other probability 
distribution involving more overlaps e.g.  $P^{12,23,34,41}(q_{12},q_{23},q_{34},q_{41})$
are uniquely determined in terms of the only independent function $P(q)$.  

The argument
is quite similar to the previous one.
Ultrametricity implies that
\begin{eqnarray}
P^{12,23,34,41}(q_{12},q_{23},q_{34},q_{41})=\\
\theta(q_{12}-q_{13})\theta(q_{23}-q_{34})\delta(q_{34}-q_{14})A(q_{12},q_{23},q_{34})
+\mbox{ permutations } +\mbox{ delta funtions}.
\end{eqnarray}
On the other hand 
\be
\int d q_{41} P^{12,23,34,41}(q_{12},q_{23},q_{34},q_{41})= 
P^{12,23,34}(q_{12},q_{23},q_{34})
\ee
and the r.h.s is known from replica equivalence.

Generally speaking ultrametricity implies that the probability distribution of the overlaps of $m$ 
replicas, which is a priori a function of $m(m-1)/2$ variables, depends only on $m-1$ variables and 
therefore, using the relations coming from the replica equivalence, is determined by the probability 
distribution with a smaller number of replicas.

Given the function $P(q)$ there is one and only one ultrametric set of probabilities satisfying the 
principle of replica equivalence.  We shall see in the next section how the ultrametric distribution 
may be characterized using a variational principle.

\section{A variational approach}
\subsection{The variational equations}
We now apply this formalism to the study of the infinite range Sherrington Kirkpatrick model near 
$T_c$.

The model is defined as follows:
\be
H=\frac12\sum_{i,k=1,N}J_{i,k}\si_{i}\si_{k},
\ee
where the $J$'s are random Gaussian variables with zero average and variance $N^{-1/2}$.

In the replica approach  the free energy density can be obtaining by finding the 
{\sl minimum} with respect to the matrix $Q$ the function $F(Q)$ defined as:
\be 
F(Q)=
\frac{\beta^{2}}{2n} \Tr Q^{2}-\frac1n \ln(\sum_{\si_{a}} \exp( \sum_{a,b}Q_{a,b}\si_{a}\si_{b})).
\ee 
The definition of {\sl minimum} is rather tricky: we say that $F(Q)$ has a minimum if its Hessian
\be
{\cal H}_{ab,cd}= {\partial^{2} F\over \partial Q_{ab}\partial Q_{cd}}
\ee
has non negative eigenvalues.  This condition may result as a minimum or as a maximum condition as 
function of the parameters on which the matrix $Q$ depends.

Near the critical temperature, where $Q$ is small, $F(Q)$ can be approximated (neglecting terms 
which do not play a crucial role) as:
\be
F(Q)=\frac1n(-\frac{\tau}2 \Tr(Q^2) +\frac13 \Tr(Q^3) -\frac{y}{4} \sum_{ab}Q^4_{ab} )\ .\label{FREE}
\ee

Let us concentrate our attention on the function $F(Q)$.  Its value is given by
\begin{equation}
W[\cP]=\int dq_{12}P(q_{12})(\frac{\tau}2 q_{12}^2+ \frac{y}{4} q_{12}^4)-
\frac13 \int dq_{12} dq_{13} dq_{23}
P^{12,23,31} (q_{12}, q_{13}, q_{23}) q_{12} q_{13} q_{23},
\end{equation}
where we indicate with $\cP$ the whole set of $P$ functions, which can be computed in terms of the 
matrix $Q$.  Near $T_c$, where we can approximate the $F(Q)$ with a polynomial, only a 
finite number of $P$ functions are relevant.

Now there is a apparently strange phenomenon.  We have remarked that the 
matrix $Q$ is a compact form for coding the the probability functional $\cP$ and it does not contains 
extra information.  We would thus expect that the equations we get from the condition
\be
{\partial W \over \partial Q_{a,b}}=0 \label{Q}
\ee
are the same that those obtained from the condition
\be
{\delta W \over \delta \cP} =0 \label{P}.
\ee

This is not the case.
Equation (\ref{Q}) implies that
\be
\tau Q_{a,b} +y Q_{a,b}^3 =\sum_c Q_{a,c}Q_{c,b}.\label{P1}
\ee
This equation correspond to an infinite set of equations for the probability distribution.
For example if  we multiply it by $ Q_{a,b}^k$ for arbitrary $k$ we arrive to the equation
\be
P(q)(\tau q+ y q^3)=\int ds dt P^{12,23,31}(q,s,t) s t \label{P2}
\ee
If we square it  and we multiply again  by $ Q_{a,b}^k$ for arbitrary $k$ we arrive to the equation
\bea
P(q)(\tau q+ y q^3)^2=\int ds dt P^{12,23,31}(q,s,t) s^2 t^2 \label{P3}\\
+\int ds dt du dv P^{12,23,31,24,41}(q,s,t,u,v) s t u v.
\eea
On the other hand the equation (\ref{P}) depends only on the probabilities with one or three 
overlaps.  Different forms of the matrix $Q$ which gives the same probability distributions with one 
and three overlaps, will give the same value of the free energy.

In the usual approach we do not see any difference among equations (\ref{P}) and (\ref{Q}).  The key 
point is separability (discussed in the next section) which is satisfied in the usual approach.  
Separability implies that the equation (\ref{P3}) is a consequence of equation (\ref{P2}).  In the 
usual approach the construction of the matrix $Q$ has been done using symmetry principles.  The 
matrix $Q$ depends on a sets of parameters $q$ and the equation (\ref {Q}) is equivalent to
\be
{\partial F \over \partial q}=0. \label {q}
\ee
If we work from a general point of view, the two equations looks rather different.  It is clear that
(\ref{Q}) contains more information than (\ref {P}) if we stay in case in which the separability 
conditions does not hold, however the last one should be enough to 
compute the free energy in any case.

Let us try to 
spell out the consequences of eq.  (\ref{P}).  It is clear that we cannot change all the $P$ in an 
independent way.  
\begin{itemize}
\item The probability with more than one overlaps are obviously related to that with
less overlap. 
\item We have the constraint that the probabilities are not negative.
\item On top of all these
requirements Guerra's relations  should be satisfied.
\end{itemize}
 We  propose the following
variational principle which should equivalent the the usual one: the free energy and the correct 
probability distributions are obtained by finding the maximum of  the functional $W[\cP]$ with respect 
to the function $P(q)$ and finding the minimum with respect to the other function $P^{12,23,31}$. This 
principle is applied 
in the region where all 
the previous constraints are satisfied.  In other words we suppose the probability are non negative 
and that
\be 
\int dq_{1,2} P^{12,23,31} (q_{1,2}, q_{1,3}, q_{2,3}) =P^{12,23} (q_{1,2}, q_{2,3})= \frac12 
P(q_{1,2})P(q_{1,3})+
\frac12 P(q_{1,2}) \delta(q_{1,2}-q_{1,3}) \label{MAGIC}
\ee
The equation (\ref{MAGIC}) is extremely important because it tell us we cannot change the function 
$P^{12,23,31}(q_{1,2}, q_{1,3}, q_{2,3})$ in an arbitrary way.

It may looks strange to maximize with respect to one parameter and to minimize with respect to an 
other, but this is quite a common situation in the replica approach \cite{MPV}.

We will argue now that this  
variational principle gives the usual ultrametric solution. It is clear that
if ultrametricity is assumed, we recover the older approach, where the free energy is maximized
with respect to $P(q)$. The only think we have to prove is that the minimization with respect to 
$P^{12,23,31}$ at fixed $P(q)$ implies ultrametricity.
In order to see how it could be done let us consider two simple case.
\subsection{An example: the overlap has two possible values}
In the first case the overlap can take two values $q_0$ with probability $p_0$ and $q_1$ with 
probability $p_1$ (we suppose $q_{0}<q_{1}$).  For the three overlaps we have four possibilities 
(all three are equal to $q_1$, two to $q_1$ and one to $q_0$
\ldots) and the corresponding probabilities are denoted $p_{111},p_{110},p_{100},p_{000}$ with 
obvious notation. The free energy is thus
\be
p_0(\frac{\tau}2 q_0+\frac{y}{4}  q_0^3)+p_1(\frac{\tau}2 q_1+ \frac{y}{4} q_1^3)
-\frac13( p_{111}q_1^3+ 3 p_{110}q_1^2 q_0+3 p_{100}q_1 q_0^2+p_{000}q_0^3)
\ee
We have that:
\bea
p_{111}+p_{110}=p_{11}=\frac12 (p_{1}^{2}+p_{1}), \\
p_{110}+p_{100}=p_{10}=\frac12 p_{1}p_{0}, \\
p_{100}+p_{000}=p_{10}= \frac12 (p_{0}^{2}+p_{0}).  \nonumber
\eea
The second equalities of each line follow from Guerra's relations.  The previous relations can be 
written as
\bea 
p_{111}=\frac12 (p_1^2+p_1) -u,\\
p_{110}=u, \\
p_{100}=\frac12 p_1 p_0 -u, \nonumber \\
 p_{000}= p_0^2  -u. \nonumber
\eea

Let us concentrate our attention on the dependence of the free energy on $u$.
We find that 
\be
F(u)=F(0)+ u (q_1-q_0)^3.
\ee
The coefficient of $u$ is positive and the minimum of the free energy with respect to $u$ is located 
at the minimum value of $u$, i.e.  $u=0$.  It is crucial to note that $u=0$ is just the condition 
which follows from the ultrametricity condition.  Using the result $u=0$ the remaining parameters 
and be computed as usual.
\subsection{An other example: the overlap has three possible values}

The same phenomenon happens if we consider a more complicated situation: three of more values of
the overlap.  Let us study what happens for three possible values of the overlap.  After some 
algebra one finds
\bea
p_{222}=\frac12(p_2+p_2^2) -a -b, \ \ \ 
p_{221}=a, \ \ \ p_{220}=b, \\
p_{211}= \frac12 p_1 p_2 -a -c,\ \ 
p_{210}=c, \ \ p_{200}= \frac12 p_2 p_0 -b -c,\\
p_{111}=\frac12(2 p_1-2 p_1 p_2-p_0p_1)+a+c-d, \ \ p_{110}=d, \nonumber \\
p_{100}= \frac12 p_1 p_0-c . 
\ \ p_{000}=p_0^2-d+b+2c, \nonumber
\eea
One finds finally the free energy depends on $a$, $b$, $c$ and $d$ as follows
\be
F(a,b,c,d)=F(0,0,0,0)+a(q_2-q_1)^3 +b(q_2-q_0)^3+d(q_1-q_0)^3
+c (q_1-q_0)^2(3q_2-q_1-2q_0).
\ee
All the coefficient are positive definite in the region where $q_0<q_1<q_2$.  The minimum of the free 
energy is reached at $a=b=c=d=0$, and the ultrametricity conditions are satisfied.

In general the distribution probability of three replicas can be written as a form which
satisfies the ultrametricity relations plus a reminder.  The dependence of 
the free energy on the reminder is linear and therefore its minimum is located just at the boundary 
of the allowed region.  The sign structure is crucial to obtain that the boundary which correspond 
to a minimum is located at the point at which the ultrametricity violations have the minimum 
possible value (i.e.  zero).  Once that the ultrametricity structure is recovered, the expression 
for the free energy is the usual one.

We are thus in the strange situation: in order to obtain the ultrametric solution we must maximize 
the free energy with respect to the function $P(q)$, but minimize it with respect to the other 
parameters which describe the violations of the ultrametricity.

On the other hand, the other possibility, the one which correspond to maximize everything, produces a 
maximal violation of the ultrametricity which would give rather nonsensical results.  Indeed in the 
first case we would get $p_{100}=0$ and $p_{110}>0$, which would clash with the interpretation in 
terms of states.  If $q_{1,2}=q_{1}$ 1 and 2 belongs to the same state and if $q_{1,3}=q_{1}$ 1 and 
3 belongs to the same; by transitivity 1 and 3 must belong to the same state and consequently 
$q_{13}=q_{1}$.

It is also not clear if the non ultrametric solution does have a replica 
interpretation, i.e.  there is an associated matrix $Q$.

It is likely worthwhile to look again to the equations coming from the cavity approach using as far a 
possible a general approach in which the ultrametricity is not assumed in order to see if there 
exists a non-ultrametric solution.  Equation (\ref{P}) or equivalently its consequences eqs.  
(\ref{P1},\ref{P2}) should be derivable in the framework of the cavity approach.  It may be possible 
that subleading terms must be considered in order to obtain interesting results.

\section{Other algebraic structures}
We have seen in the previous section that ultrametricity is an extremely powerful constraint and 
that there is only one ultrametric probability distribution at fixed ($P(q)$).  In this section we 
explore other algebraic properties that are present in the usual solution.  The aim is to 
characterize as far as possible the probability distribution in order to find how much space there 
is for non ultrametric solutions, which should still have good properties.

The properties which we are going to consider are translational invariance in 
replica space and separability.
\subsection{Translational invariance}
Replica equivalence implies that all replicas are equivalent. The simplest way to implement it is to
assume that there a group $G$  (a subgroup of the group of permutation of $n$ 
elements) which leaves the matrix $Q$ invariant and acts transitively on the space of indices (i.e.  
for any pairs $a,b$ there is element $g$ of the group such that $g \cdot a =b$) \cite{BL}.

The simplest group we can think of, as suggested by Kondor \cite{FRAN}, is the translational group, 
i.e.  the group generated by the operation $a \to a+1$ (all sums and differences are done modulo 
$n$, in this section).  It was later realized by Sourlas \cite{SOU,PASOU} that by  reshuffling  
the indices the usual form of matrix $Q$ in presence of replica symmetry breaking can be put in such 
a way to be translational invariant.  In this new base we have
\begin{equation}
Q_{a,b}=q(a-b).
\end{equation}

Translational invariance is quite useful at the technical level: the matrix product becomes now a 
convolution which can be easily done by Fourier transform \cite{PASOU,FOU}.  Using this fact, if we 
call $s$ the variable of the Fourier transform, one finds that there exists a function $F(q,s)$ such 
that
\bea
\int d\mu(s) F(q_1,s) F(q_2,s)= P(q_1)\delta(q_1-q_2),\\
\int d\mu(s) F(q_1,s) F(q_2,s) F(q_3,s)= P^{12,23,31}(q_1,q_2,q_3).
\eea

Similar relations can be found for probabilities involving an high number of $q$ (i.e for 
$P^{12,23,34,41}$, $P^{12,23,34,45,51}$ and so on.  These relations are not on the same standing of 
the ones derived in the previous sections because they do not allow us to write algebraic equations 
involving a finite set of probability distributions; they however strongly constrain the set of 
allowed probability distributions.

At this stage it is not clear how to exploit  the consequences of translational invariance on the
probability distribution.  It seems reasonable to assume that not all the 
probability distributions which satisfy replica equivalence can be generated by a matrix $Q_{a,b}$ 
which is translational invariant, but it is no so easy to find a counterexample.  It is not 
clear if the alternative scheme introduced in \cite{BL} can be extended in such a way to 
have a probability interpretation.

An answer to these 
questions would be interesting, because it would help us to elucidate the origine of the rather 
mysterious translation invariance in replica space.
\subsection{Separability}
Separability (or non degeneracy) correspond to the following algebraic statement.
Let us consider  all the matrices which can be generated from the matrix $Q$ in a permutational 
covariant fashion. Some example are
\be
Q_{ab}^k, \ \ \sum_c Q_{ac}Q_{cb}, \ \ \sum_{c,d} Q_{ac}Q_{ad}Q_{cd}Q_{cb} Q_{db}.\label{SET}
\ee
Separability states that if we take two pair of indices ($ab$ and $cd$), we have that
\be
Q_{ab}=Q_{cd} \longrightarrow M_{ab}=M_{cd}\label{SEPA}
\ee
where $M$ is a generic matrix of the set generated by the rules (\ref{SET}).
In other words pairs of indices which have different properties have a different values of $Q$.

In the usual approach when replica symmetry is broken there is subgroup of the group of permutations 
that commutes with the matrix $Q$. Let us consider the orbits in the space of pairs of indices.  It 
evident that the values of the elements of the matrix $Q$ and of any matrix derived using the rules 
are 
constant of the orbits. If we assume that different orbits have different value of
the matrix $Q$ we obtain the condition (\ref{SEPA}).  

The separability condition is extremely powerful in determining the expectation values of higher order 
moments of the probability distribution.

 Let us study a simple example and let us consider a matrix
$M$ constructed with the rules (\ref{SET}). It is evident that if
\be
\sum_b Q_{a,b}^k M_{a,b}= \int dq P(q) M(q)q ^k,\ \ \sum_b Q_{a,b}^k R_{a,b}= \int dq P(q) R(q)q ^k
\ee,
we have  that
\be
\sum_b Q_{a,b}^k M_{a,b}R_{a,b} = \int dq P(q) M(q) R(q) q^k.
\ee
If we apply the previous formula to the case where $M$ and $R$ have the form
\be
M_{ab}=\sum_c Q_{ac}^{k_{1}} Q_{cb}^{k_{2}}, R_{ab}=\sum_c Q_{ac}^{k_{3}} Q_{cb}^{k_{4}}
\ee
we find the rather surprising formula
\bea
3P^{12,13,32,14,42}(q,q_1,q_2,q_3,q_4) =
\delta(q_1-q_3)\delta(q_2-q_4) 
P^{12,23,31}(q,q_1,q_2)\\
+2 {P^{12,23,31}(q,q_1,q_2)P^{12,23,31}(q,q_3,q_4) \over P(q)}.\label{WOH}
\eea
Similar results can be obtained for other probability distributions.

Equation (\ref{WOH}) is particular interesting because integrating over $q$ it implies that
\begin{eqnarray}
3P^{13,32,14,42}(q_1,q_2,q_3,q_4) =
\delta(q_1-q_3)\delta(q_2-q_4) 
(P(q_1)P(q_2) +\delta(q_{1}-q_{2})P(q_{2}))\\
+2\int dq{ P^{12,23,31}(q,q_1,q_2)P^{12,23,31}(q,q_3,q_4) \over P(q)}.\label{NEXT}
\end{eqnarray}
This 
equation can also written in more suggestive form as
\begin{eqnarray}
3P^{13,32,14,42}(q_1,q_2,q_3,q_4) =
\delta(q_1-q_3)\delta(q_2-q_4) 
(P(q_1)P(q_2) +\delta(q_{1}-q_{2})P(q_{2}))\\
+2\int dq{ P^{12,23,31}(q,q_1,q_2)P^{23,31,12}(q_3,q_4|q) },
\end{eqnarray}
where $P^{23,31,12}(q_3,q_4|q)$ is a conditional probability.

Now the l.h.s of eq.  (\ref{NEXT})is by definition invariant under cyclic permutation of the $q$'s, 
while the r.h.s in not invariant for a generic choice of the function $P^{12,23,31}$.  A boring 
computation is needed to verify that if we insert the ultrametric result of the previous section, we 
find a symmetric result (as we should) for $P^{13,32,14,42}$.

It not clear which is the most general form of the of the probability $P^{12,23,31}$, which gives is 
compatible with separability.  We have checked what happens in the case of two or three possible 
values of the overlaps, discussed in the previous section.
\begin{itemize}
\item
In the case of two possible values of the overlap, the only non trivial relation is
\be
P^{13,32,14,42}(q_{0},q_{1},q_{0},q_{1})=P^{13,32,14,42}(q_{1},q_{1},q_{0},q_{0}).
\ee
An explicit computation shows that this relation is identically satisfied.

\item
In the case of two possible values of the same relation would is not identically satisfied,
however
it is satisfied for $a=b=c$ with arbitrary $d$.
\end{itemize}

The previous examples shows that separability does not imply ultrametricity in the general case, but 
impose rather strong constraints on the distributions functions $P$.

This relations coming from separability are quite 
powerful in imposing further constraint on the probability distribution beyond replica equivalence.  
Their origine can be well understood in an algebraic framework, however it is not clear if they can 
be derived in a more general setting.  It would be extremely interesting to check if they are 
satisfied in numerical simulations.

\section*{Acknowledgments} It is a
pleasure to thank Francesco Guerra and Imre Kondor for illuminating discussions.  I am also grateful 
to M.  Aizenman and F.  Contucci for communicating to me (prior to publications) their work
\cite{AC}, which strongly overlaps with this one and for further correspondence.

\end{document}